\RequirePackage{lineno} 
\documentclass[prd,onecolumn,groupedaddress,superscriptaddress,amsmath,amssymb]{revtex4}

\usepackage{graphicx}
\usepackage{dcolumn}
\usepackage{bm}
\usepackage{amssymb}
\usepackage{enumerate}
\usepackage{lineno}

\newcommand\g{\gamma}
\newcommand\ee{e^+e^-}
\newcommand\pp{\pi^+ \pi^-}
\newcommand\mm{\mu^+ \mu^-}

\newcommand\aee{A' \to e^+e^-}
\newcommand\xdecay{X \rightarrow e^+  e^-}

\newcommand\ma{m_{A'}}
\newcommand\pair{e^+e^-}

\newcommand\Na{{N}_{A'}}

\newcommand\kosdec{K^0_S \to \pi^0 \pi^0; \pi^0 \to \g \ee}
\newcommand\kospio{K^0_S \to \pi^0 \pi^0}

\def\address{\@ifstar{\address@star}%
  {\@ifnextchar[{\address@optarg}{\address@noptarg}}}

\begin{document}



\title{Improved limits on a hypothetical $X(16.7)$ boson and a dark photon decaying into $\pair$ pairs}

\affiliation{\it Universit\"at Bonn, Helmholtz-Institut f\"ur Strahlen-und Kernphysik, 53115 Bonn, Germany} 
\affiliation{\it Joint Institute for Nuclear Research, 141980 Dubna, Russia}
\affiliation{\it Technische Universit\"at M\"unchen, Physik  Department, 85748 Garching, Germany}
\affiliation{\it CERN, European Organization for Nuclear Research, CH-1211 Geneva, Switzerland}
\affiliation{\it University of Illinois at Urbana Champaign, Urbana, 61801-3080 Illinois, USA}
\affiliation{\it UCL Departement of Physics and Astronomy, University College London, Gower St. London WC1E 6BT, United Kingdom}
\affiliation{\it Institute for Nuclear Research, 117312 Moscow, Russia}
\affiliation{\it P.N. Lebedev Physical Institute, Moscow, Russia, 119 991 Moscow, Russia}
\affiliation{\it Skobeltsyn Institute of Nuclear Physics, Lomonosov Moscow State University, 119991  Moscow, Russia}
\affiliation{\it Physics Department, University of Patras, 265 04 Patras, Greece} 
\affiliation{\it State Scientific Center of the Russian Federation Institute for High Energy Physics of National Research Center 'Kurchatov Institute' (IHEP), 142281 Protvino, Russia}
\affiliation{\it Departamento de Ciencias F\'{i}sicas, Universidad Andres Bello, Sazi\'{e} 2212, Piso 7, Santiago, Chile}
\affiliation{\it Tomsk State Pedagogical University, 634061 Tomsk, Russia}
\affiliation{\it Universidad T\'{e}cnica Federico Santa Mar\'{i}a, 2390123 Valpara\'{i}so, Chile}
\affiliation{\it ETH Z\"urich, Institute for Particle Physics and Astrophysics, CH-8093 Z\"urich, Switzerland}
\author{D.~Banerjee}\affiliation{\it CERN, European Organization for Nuclear Research, CH-1211 Geneva, Switzerland}\affiliation{\it University of Illinois at Urbana Champaign, Urbana, 61801-3080 Illinois, USA}
\author{J.~Bernhard}\affiliation{\it CERN, European Organization for Nuclear Research, CH-1211 Geneva, Switzerland}
\author{V.~E.~Burtsev}\affiliation{\it Joint Institute for Nuclear Research, 141980 Dubna, Russia}
\author{A.~G.~Chumakov}\affiliation{\it Tomsk State Pedagogical University, 634061 Tomsk, Russia}
\author{D.~Cooke}\affiliation{\it UCL Departement of Physics and Astronomy, University College London, Gower St. London WC1E 6BT, United Kingdom}
\author{P.~Crivelli}\affiliation{\it ETH Z\"urich, Institute for Particle Physics and Astrophysics, CH-8093 Z\"urich, Switzerland}
\author{E.~Depero}\affiliation{\it ETH Z\"urich, Institute for Particle Physics and Astrophysics, CH-8093 Z\"urich, Switzerland}
\author{A.~V.~Dermenev}\affiliation{\it Institute for Nuclear Research, 117312 Moscow, Russia}
\author{S.~V.~Donskov}\affiliation{\it State Scientific Center of the Russian Federation Institute for High Energy Physics of National Research Center 'Kurchatov Institute' (IHEP), 142281 Protvino, Russia}
\author{R.~R.~Dusaev}\affiliation{\it Tomsk State Pedagogical University, 634061 Tomsk, Russia}
\author{T.~Enik}\affiliation{\it  Joint Institute for Nuclear Research, 141980 Dubna, Russia}
\author{N.~Charitonidis}\affiliation{\it CERN, European Organization for Nuclear Research, CH-1211 Geneva, Switzerland}
\author{A.~Feshchenko}\affiliation{\it  Joint Institute for Nuclear Research, 141980 Dubna, Russia}
\author{V.~N.~Frolov}\affiliation{\it  Joint Institute for Nuclear Research, 141980 Dubna, Russia}
\author{A.~Gardikiotis}\affiliation{\it Physics Department, University of Patras, 265 04 Patras, Greece}
\author{S.~G.~Gerassimov }\affiliation{\it P.N. Lebedev Physical Institute, Moscow, Russia, 119 991 Moscow, Russia}\affiliation{\it Technische Universit\"at M\"unchen, Physik  Department, 85748 Garching, Germany}
\author{S.~N.~Gninenko}\affiliation{\it Institute for Nuclear Research, 117312 Moscow, Russia}
\author{M.~H\"osgen}\affiliation{\it Universit\"at Bonn, Helmholtz-Institut f\"ur Strahlen-und Kernphysik, 53115 Bonn, Germany}
\author{M.~Jeckel}\affiliation{\it CERN, European Organization for Nuclear Research, CH-1211 Geneva, Switzerland}
\author{V.~A.~Kachanov}\affiliation{\it State Scientific Center of the Russian Federation Institute for High Energy Physics of National Research Center 'Kurchatov Institute' (IHEP), 142281 Protvino, Russia}
\author{A.~E.~Karneyeu}\affiliation{\it Institute for Nuclear Research, 117312 Moscow, Russia}
\author{G.~Kekelidze}\affiliation{\it  Joint Institute for Nuclear Research, 141980 Dubna, Russia}
\author{B.~Ketzer}\affiliation{\it Universit\"at Bonn, Helmholtz-Institut f\"ur Strahlen-und Kernphysik, 53115 Bonn, Germany}
\author{D.~V.~Kirpichnikov}\affiliation{\it Institute for Nuclear Research, 117312 Moscow, Russia}
\author{M.~M.~Kirsanov}\affiliation{\it Institute for Nuclear Research, 117312 Moscow, Russia}
\author{V.~N.~Kolosov}\affiliation{\it State Scientific Center of the Russian Federation Institute for High Energy Physics of National Research Center 'Kurchatov Institute' (IHEP), 142281 Protvino, Russia}
\author{I.~V.~Konorov}\affiliation{\it P.N. Lebedev Physical Institute, Moscow, Russia, 119 991 Moscow, Russia} \affiliation{\it Technische Universit\"at M\"unchen, Physik  Department, 85748 Garching, Germany}
\author{S.~G.~Kovalenko}\affiliation{\it Departamento de Ciencias F\'{i}sicas, Universidad Andres Bello, Sazi\'{e} 2212, Piso 7, Santiago, Chile}
\author{V.~A.~Kramarenko}\affiliation{\it  Joint Institute for Nuclear Research, 141980 Dubna, Russia}\affiliation{\it Skobeltsyn Institute of Nuclear Physics, Lomonosov Moscow State University, 119991  Moscow, Russia}
\author{L.~V.~Kravchuk}\affiliation{\it Institute for Nuclear Research, 117312 Moscow, Russia}
\author{ N.~V.~Krasnikov}\affiliation{\it  Joint Institute for Nuclear Research, 141980 Dubna, Russia}\affiliation{\it Institute for Nuclear Research, 117312 Moscow, Russia}
\author{S.~V.~Kuleshov}\affiliation{\it Departamento de Ciencias F\'{i}sicas, Universidad Andres Bello, Sazi\'{e} 2212, Piso 7, Santiago, Chile}
\author{V.~E.~Lyubovitskij}\affiliation{\it Tomsk State Pedagogical University, 634061 Tomsk, Russia}\affiliation{\it Universidad T\'{e}cnica Federico Santa Mar\'{i}a, 2390123 Valpara\'{i}so, Chile}
\author{V.~Lysan}\affiliation{\it  Joint Institute for Nuclear Research, 141980 Dubna, Russia}
\author{V.~A.~Matveev}\affiliation{\it  Joint Institute for Nuclear Research, 141980 Dubna, Russia}
\author{Yu.~V.~Mikhailov}\affiliation{\it State Scientific Center of the Russian Federation Institute for High Energy Physics of National Research Center 'Kurchatov Institute' (IHEP), 142281 Protvino, Russia}
\author{L.~Molina Bueno}\affiliation{\it ETH Z\"urich, Institute for Particle Physics and Astrophysics, CH-8093 Z\"urich, Switzerland}
\author{D.~V.~Peshekhonov}\affiliation{\it  Joint Institute for Nuclear Research, 141980 Dubna, Russia}
\author{V.~A.~Polyakov}\affiliation{\it State Scientific Center of the Russian Federation Institute for High Energy Physics of National Research Center 'Kurchatov Institute' (IHEP), 142281 Protvino, Russia}
\author{B.~Radics}\affiliation{\it ETH Z\"urich, Institute for Particle Physics and Astrophysics, CH-8093 Z\"urich, Switzerland}
\author{R.~Rojas}\affiliation{\it Universidad T\'{e}cnica Federico Santa Mar\'{i}a, 2390123 Valpara\'{i}so, Chile}
\author{A.~Rubbia}\affiliation{\it ETH Z\"urich, Institute for Particle Physics and Astrophysics, CH-8093 Z\"urich, Switzerland}
\author{V.~D.~Samoylenko}\affiliation{\it State Scientific Center of the Russian Federation Institute for High Energy Physics of National Research Center 'Kurchatov Institute' (IHEP), 142281 Protvino, Russia}
\author{D.~Shchukin}\affiliation{\it P.N. Lebedev Physical Institute, Moscow, Russia, 119 991 Moscow, Russia}
\author{V.~O.~Tikhomirov}\affiliation{\it P.N. Lebedev Physical Institute, Moscow, Russia, 119 991 Moscow, Russia}
\author{I.~Tlisova}\affiliation{\it Institute for Nuclear Research, 117312 Moscow, Russia} 
\author{D.~A.~Tlisov}\affiliation{\it Institute for Nuclear Research, 117312 Moscow, Russia} 
\author{A.~N.~Toropin}\affiliation{\it Institute for Nuclear Research, 117312 Moscow, Russia}
\author{A.~Yu.~Trifonov}\affiliation{\it Tomsk State Pedagogical University, 634061 Tomsk, Russia}
\author{B.~I.~Vasilishin}\affiliation{\it Tomsk State Pedagogical University, 634061 Tomsk, Russia}
\author{G.~Vasquez Arenas}\affiliation{\it Universidad T\'{e}cnica Federico Santa Mar\'{i}a, 2390123 Valpara\'{i}so, Chile}
\author{P.~V.~Volkov}\affiliation{\it  Joint Institute for Nuclear Research, 141980 Dubna, Russia}\affiliation{\it Skobeltsyn Institute of Nuclear Physics, Lomonosov Moscow State University, 119991  Moscow, Russia}
\author{V.~Yu.~Volkov}\affiliation{\it Skobeltsyn Institute of Nuclear Physics, Lomonosov Moscow State University, 119991  Moscow, Russia}
\author{P.~Ulloa}\affiliation{\it Universidad T\'{e}cnica Federico Santa Mar\'{i}a, 2390123 Valpara\'{i}so, Chile}

%
%
\collaboration{The NA64 Collaboration}\noaffiliation
\vskip 0.25cm


\begin{abstract}
  The improved results on a direct search for a new $X$(16.7 MeV) boson that could explain the anomalous excess of $\ee$ pairs observed
in the decays of the excited $^8$Be$^*$ nucleus ("Berillium anomaly") are reported. The $X$ boson could be produced
in the bremsstrahlung reaction $e^- Z \to e^- Z  X$ by a high energy beam of electrons incident on the active target in the NA64 experiment
at the CERN SPS and observed through its subsequent decay into $\ee$ pair. No evidence for such decays was found from the combined analysis
of the data samples with total statistics corresponding to $8.4\times 10^{10}$ electrons on target collected in 2017 and 2018.
This allows to set the new limits on the $X-e^-$ coupling in the range $1.2 \times 10^{-4}\lesssim \epsilon_e \lesssim 6.8 \times 10^{-4}$,
excluding part of the parameter space favored by the Berillium anomaly.
The non-observation of the decay $\aee$ allows also to set the new bounds on the mixing strength of photons with dark photons ($A'$)
with a mass $\lesssim 24$ MeV.
\end{abstract}

\maketitle

 Recently, the search for new light bosons weakly coupled to SM particles was additionally inspired by the observation in the ATOMKI
experiment by Krasznahorkay et al. \cite{be8,be8-1} of a $\sim$7$\sigma$ excess of events in the invariant mass distribution
of $\pair$ pairs produced in the nuclear transitions of the excited $^8$Be$^*$ to its ground state via internal pair creation.
It was shown that this anomaly can be interpreted as the emission of a protophobic
gauge boson $X$ with a mass of 16.7 MeV decaying into $\pair$ pair \cite{feng1,feng2}.
This explanation of the anomaly was found to be consistent with the existing constraints assuming that the $X$ has
non-universal coupling to quarks, coupling to electrons in the range $2\times 10^{-4} \lesssim \epsilon_e \lesssim 1.4\times 10^{-3}$
and lifetime $10^{-14}\lesssim \tau_X \lesssim 10^{-12}$~s. It is interesting that a new boson with such relatively large
couplings to charged leptons could also resolve the so-called ($g_\mu - 2$ ) anomaly, a discrepancy between measured and predicted
values of the muon anomalous magnetic moment. This has motivated worldwide efforts towards the experimental searches,
see, e.g., Refs.~\cite{mb, nardi}, and studies of the phenomenological aspects of light vector bosons weakly coupled to quarks and leptons,
see, e.g., Refs.~\cite{jk, cheng, Zhang:2017zap, ia, liang, bart} and also earlier works of Refs.~\cite{fayet1, fayet2, fayet3, fayet4}.
The latest experimental results from the ATOMKI group show a similar excess of events at approximately the same invariant mass
in the nuclear transitions of another nucleus, $^4$He \cite{be8-2}. This further increases the importance of independent searches for
a new particle $X$.
 \par Another strong motivation to search for new light bosons decaying into $\ee$ pair comes from the
dark matter puzzle. An interesting possibility is that in addition to gravity a new force between the dark sector and
visible matter, carried by a new vector boson $A'$, called dark photon, might exist \cite{prw, pospelov}. Such $A'$ could have 
a mass $m_{A'}\lesssim 1$ GeV, associated with a spontaneously broken gauge $U(1)_D$ symmetry, and would couple to 
the Standard Model (SM) through the kinetic mixing with ordinary photon, $-\frac{1}{2}\epsilon F_{\mu\nu}A'^{\mu\nu}$, parameterized
by the mixing strength  $\epsilon \ll 1$ \cite{Okun:1982xi, Galison:1983pa, Holdom:1985ag}, for a review see, e.g., Refs.~\cite{mb, jr, report}. 

\begin{figure*}[tbh!!]
\centering
\includegraphics[width=.9\textwidth]{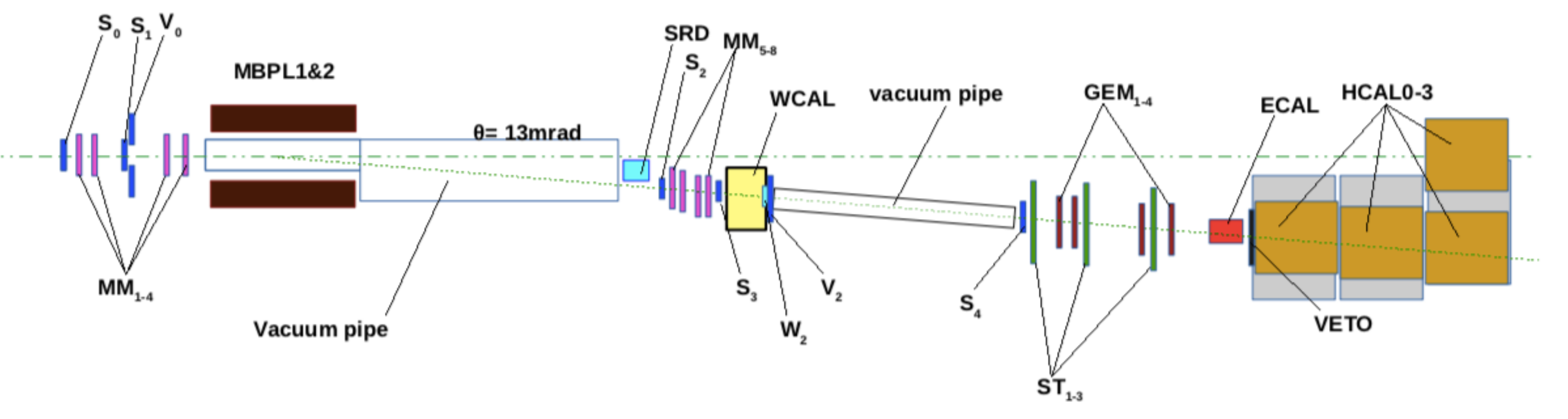}
\caption{The setup to search for $A',X\to \ee$  decays of the bremsstrahlung $A',X$ produced in the reaction
$eZ \to eZA'(X) $ of the 150 GeV electrons incident on the active WCAL target.}
\label{setup}
\end{figure*}

 A number of previous experiments, such as beam dump \cite{jdb, charm, rio, e137, konaka, bross, dav,  ath, nomad, e787, essig1, blum,
sg1, blum1, sarah1}, fixed target \cite{apex,merkel,merkel1}, collider \cite{babar, curt, babar1} and rare particle decay searches
\cite{bern, sindrum, kloe, sg2, kloe2, wasa, hades, phenix, e949, na48, pol, kloe3}, put stringent
constraints on the $\epsilon$ and mass $m_{A'}$ of such dark photons, excluding, in particular,
the parameter space region favored by the $g_\mu-2$ anomaly. However, a large range of mixing strengths 
$10^{-4} \lesssim \epsilon \lesssim 10^{-3}$ corresponding to short-lived $A'$ remains unexplored. These values of $\epsilon$
can be obtained from the loop effects of particles charged under both the dark and SM $U(1)$ interactions.
Typically 1-loop value is $\epsilon = e g_D/16\pi^2$ \cite{Holdom:1985ag}, where $g_D$ is the coupling constant of the $U(1)_D$
gauge interactions. The search for $\ee$ decays of new short-lived particles at the CERN SPS was performed by the NA64 experiment
in 2017 \cite{NA64Be2017}. We report here the improved results from the NA64 experiment obtained using the data collected
in 2018 in the new run at the CERN SPS performed after optimization of the experiment configuration and parameters.

 \par The NA64 experiment employs the optimized electron beam from the H4 beam line of the CERN SPS.
The beam delivers $\simeq 5\times 10^6~e^-$ per SPS spill of 4.8 s produced by the primary 400 GeV proton beam with an intensity
of a few 10$^{12}$ protons on target. The NA64 setup designed for the searches of $X$ bosons and $A'$ is schematically
shown in Fig.~\ref{setup}. The thin scintillation counters, $S_1$ - $S_3$ and $V_0$, are used for the beam definition, while another one,
$S_4$, is used to detect the $\pair$ pairs. The detector is equipped with a magnetic spectrometer consisting of two MBPL magnets
and a tracker with low material budget. The tracker is a set of four upstream Micromegas (MM) chambers for the incoming
$e^-$ angle selection, four GEM chambers and three straw tube planes allowing the reconstruction of the outgoing tracks \cite{Banerjee:2015eno,track}.
To enhance the electron identification the synchrotron radiation (SR) emitted by electrons is used for their efficient
tagging and for additional suppression of the initial hadron contamination in the beam $\pi/e^- \simeq 10^{-2}$ down to the level
$\simeq 10^{-6}$ \cite{srd,na64-prd}. The use of SR detectors (SRD) is important for the hadron background suppression and the corresponding
improvement of the sensitivity as compared to the previous electron beam dump searches \cite{konaka,bross}.
 The dump is an electromagnetic (EM) calorimeter WCAL made as compact as possible to maximize the sensitivity to short
lifetimes while keeping the leakage of particles at a small level. The purpose of the WCAL design was to absorb {\it not the full energy}
of the shower generated by the primary electrons, but the energy of the showers produced by secondary particles and the recoil electrons
from the primary reaction (1), which is typically significantly lower. The WCAL is assembled from the tungsten and plastic scintillator
plates with wave length shifting fiber read-out. The first five layers of the WCAL are separated from the main part (WCAL preshower). Immediately
after the WCAL there are the veto counters $W_2$ and $V_2$, several meters downstream the decay counter $S_4$ and tracking detectors.
These detectors are followed by another EM calorimeter (ECAL), which is a matrix of $6\times 6$ shashlik-type lead - plastic scintillator sandwich
modules \cite{na64-prd}. The ECAL is 40 radiation lengths (X0) with the first 4 X0 serving as a preshower subdetector.
Downstream the ECAL the detector is equipped with a high-efficiency counter VETO and a thick hadron calorimeter (HCAL) \cite{na64-prd}
used as a hadron veto and muon identificator.

 The events are collected with a hardware trigger requiring in-time energy deposition in $S_1$ - $S_3$, no energy deposition in $V_0$ and
$E_{WCAL} \lesssim 0.7 \times E_{beam}$. The latter requirement was not used in the runs used for calibration (calibration beams).

 In order to increase the sensitivity to short-lived X bosons (higher $\epsilon$) the following optimization steps
were performed for the 2018 run: (i) Beam energy increased to 150 GeV; (ii) Thinner counter $W_2$ was installed
immediately after the last tungsten plate inside the WCAL box; (iii) more track detectors installed between WCAL and ECAL.
In addition, the vacuum pipe was installed immediately after the WCAL, and the distance between the WCAL and ECAL was increased.
These changes would allow to perform the full track and vertex reconstruction if the $\ee$ pair energy is not very high
as immediate additional checks in case of signal observation.

 To choose selection criteria, for the calculation of efficiencies and for background estimation the package based on
Geant4 \cite{Geant4-2002, Geant4-2006} for the detailed full simulation of the experiment is developed. It contains the subpackage
for the simulation of various types of dark matter particles based on the exact tree-level calculation of cross sections \cite{DMsimulation}.

 \par The method of the search for $\aee$ (or $\xdecay$) decays is described in \cite{Gninenko:2013rka, Andreas:2013lya, gkkk1, DMsimulation}.
The $A'$ can be produced via the coupling to electrons in the scattering of high-energy electrons off nuclei of the
active WCAL target-dump. Its production is followed by the decay into $\ee$ pairs:

\begin{equation}
e^- + Z \to e^- + Z + A'(X)   ;~ A'(X)\to \ee \,.
\label{ea}
\end{equation}
 
The $A'$ would penetrate the dump and the veto counter without interactions and then decay in flight
into an $\ee$ pair in the decay volume. A fraction~($f$) of the primary electron energy $E_1 = f E_0$ is deposited
in the WCAL by the recoil electron from the reaction \eqref{ea}. The WCAL serves thus as a dump to absorb the EM shower
from the recoil and other secondary particles produced before the $A'$ production.
The other, typically bigger part of the primary electron energy $E_2 = (1-f)E_0$, is transferred through the dump by the~$A'$ and deposited
in the second calorimeter ECAL via the $A'(X)\to \ee$ decay in flight, as shown in Fig.~\ref{setup}.

 The occurrence of $A'(X)$ produced in the $e^- Z $ interactions and $\aee$ decays would appear as an excess of events with two EM-like showers
in the setup: one shower in the WCAL and another one in the ECAL, with the total energy $E_{tot} = E_{WCAL} +E_{ECAL}$
compatible with the beam energy ($E_0$), above those expected from the background sources.

 \par The candidate events were selected with the following criteria:
(i) Small energy in the veto counter ($W_2$ in 2018), well below one $MIP$ (most probable energy deposition
of a minimum ionizing particle). The concrete cut was slightly different for different periods, it was optimized taking into account
the energy resolution, the electronic noise and the pileup effects in the counter;
(ii) The signal in the decay counter $S_4$ is consistent with two $MIP$s;
(iii) The sum of energies deposited in the WCAL+ECAL is equal to the beam energy within the boundaries determined by the energy
resolution of these detectors. At least 25 GeV should be deposited in the ECAL;
(iv) The shower in the WCAL should start to develop within a few first $X_0$, which is ensured by the WCAL preshower energy cut;
(v) The cell with maximal energy deposition in the ECAL should be (3,3): the cell on the axis of the beam bent by the magnets;
(vi) The longitudinal and lateral shape of the shower in the ECAL are consistent with a single EM one. The longitudinal shape
is checked by the cut on the energy deposition in the ECAL preshower. Checking the lateral shower shape
does not decrease the efficiency for signal events because the distance between $e^-$ and $e^+$ in the ECAL is significantly smaller
than the ECAL cell size. Finally, the rejection of events with hadrons in the final state was based on the energy deposition in the
VETO and HCAL.

 As in the previous analyses \cite{na64-prl, na64-prd}, in order to check efficiencies and the reliability of the MC simulations,
we selected a clean sample of $\simeq 10^5$ $\mu^+ \mu^-$ events with $E_{WCAL} < 0.6 \times E_{beam}$ from the QED muon pair
production in the dump (dimuons). This rare process is dominated by the reaction $e^- Z \to e^- Z \gamma; \gamma \to \mu^+ \mu^-$
of a photon conversion into muon pair on a dump nucleus. We performed a number of comparisons between these
events and the corresponding MC simulated sample and applied the estimated efficiency corrections to the MC events.

\begin{figure}
\centering
\includegraphics[width=0.45\textwidth]{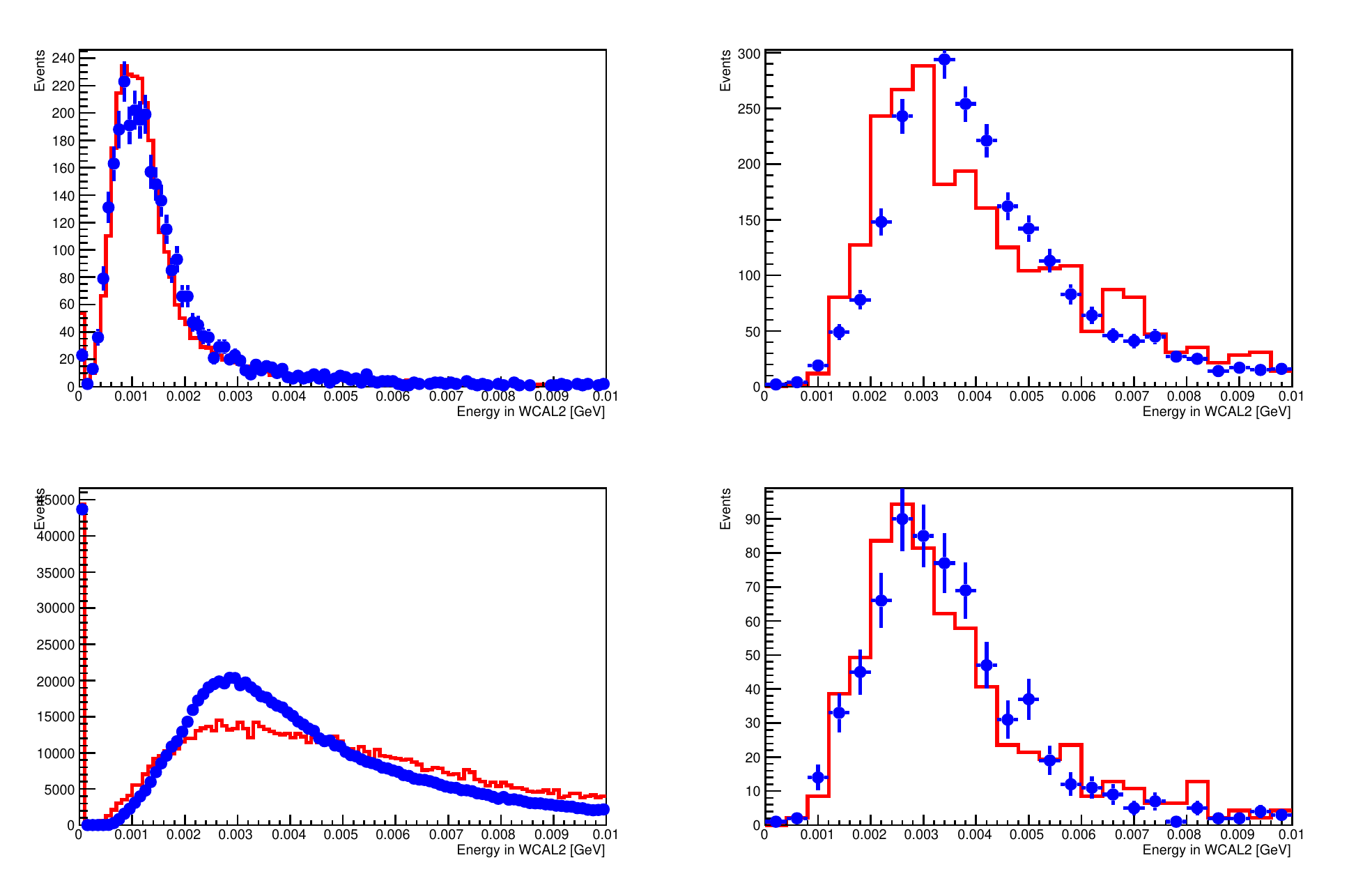}
\caption{The comparison of various distributions in the $W_2$ counter with MC predictions. Left upper plot: for muons
selected in the hadron calibration beam. Right upper plot: for dimuon events with the standard cut $E_{WCAL} < 0.6 \times E_{beam}$.
Right lower plot: for dimuon events with the cut $E_{WCAL} < 0.33 \times E_{beam}$. Left lower plot: for the electron
calibration beam.}
\label{W2checks}
\end{figure}

 The counter $W_2$ is very important for this analysis. It is made using the same technology as for the tiles of the WCAL and
installed inside the WCAL box to be as close to the possible place of $A'$ creation as possible. We paid special attention
to check that it works correctly and to make the MC simulation of this counter as close to the real data as possible.
In the simulation we took into account the following effects:
\begin{itemize}
\item fluctuations of the number of photoelectrons from the photocathode
\item pulse reconstruction threshold curve for the counter below 0.8 MIP
\item small cross-talk between the WCAL and W2 signals
\item uncertainties of the $W_2$ pulse reconstruction due to readout electronic noise and pileup effects
\end{itemize} 
The cross-talk between the neighboring WCAL and W2 signals include contributions from the light cross-talk and the electronic
cross-talk between the two channels. The average cross-talk value was assumed to be proportional to the energy deposition in the WCAL.

in Fig.~\ref{W2checks} the comparison of the MC simulation with data for selected muons in the hadron beam and for the electron beam
with several different selections is shown. There is some remaining disagreement for the electron calibration beam and for dimuons.
However, the agreement for dimuons becomes better for smaller energy in the WCAL, i.e. for the conditions that we would have in
signal events. For reliability we also estimated the systematic errors of the signal efficiency due to $W_2$ by changing the $W_2$
threshold (30\% up and down) and comparing the signal efficiencies. The systematic error calculated this way is 10\%. It was used
in the final statistical analysis together with other systematic errors.

\begin{figure}
\centering
\includegraphics[width=0.45\textwidth]{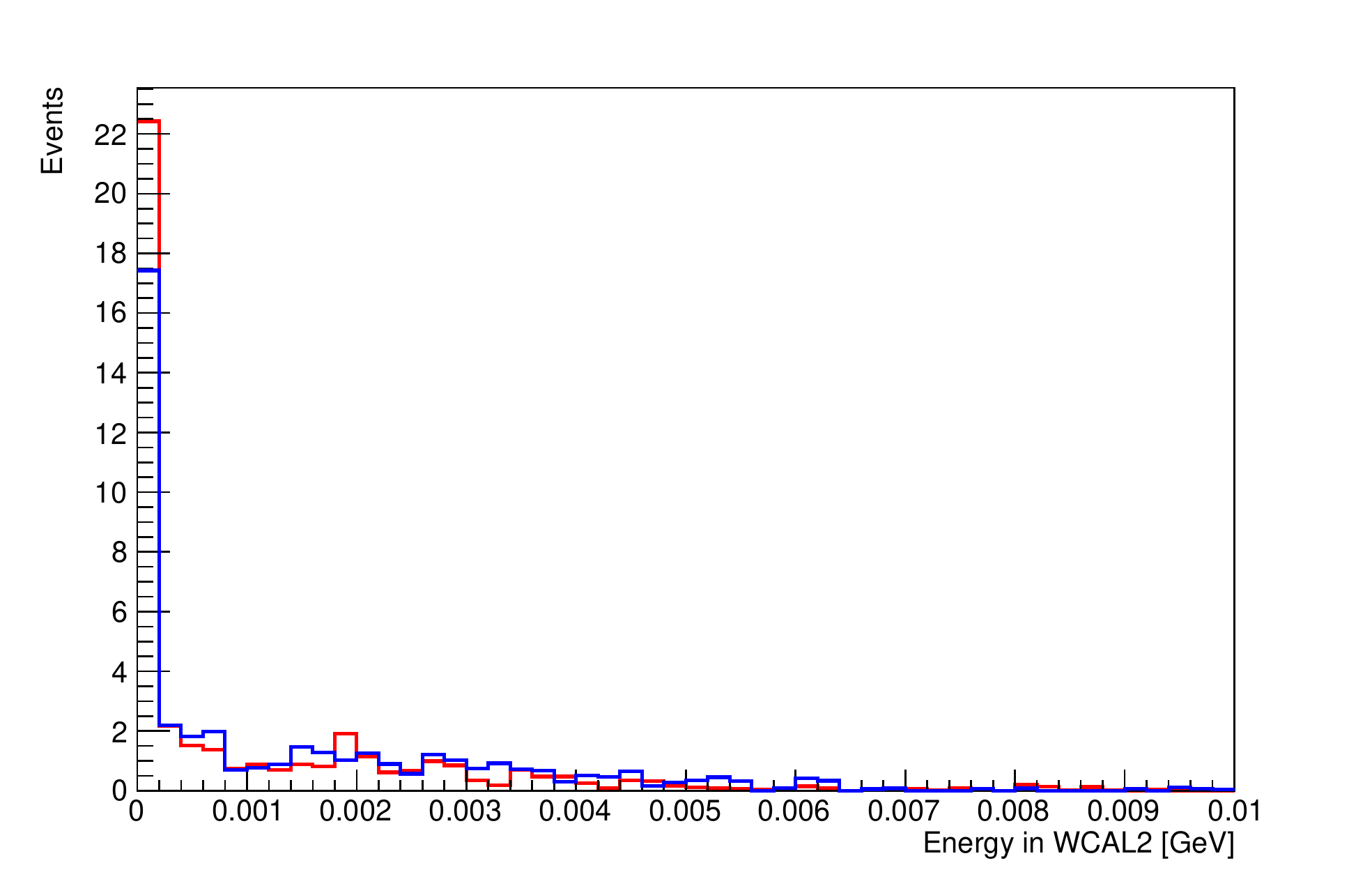}
\caption{The expected energy in the $W_2$ counter from the signal events. Blue histogram: $\epsilon = 0.0003$.
Red histogram: $\epsilon = 0.0006$. The events are histogrammed with the weight corresponding to the probability for
the $A'$ to decay after the last tungsten plate of the WCAL.}
\label{W2signal}
\end{figure}

 The energy deposition in $W_2$ expected for the detectable signal events (with $A'$ decays after the last tungsten plate) is shown
in Fig.~\ref{W2signal}. It is significantly smaller than for the electrons from the primary beam (Fig.~\ref{W2checks} lower left
plot). It is also smaller for the bigger value of $\epsilon$ since the short-lived $A'$ should have higher energy for the same
probability to decay after the WCAL tungsten plates, which means smaller energy of the recoil electron (shorter shower).

\begin{figure}
\centering
\includegraphics[width=0.45\textwidth]{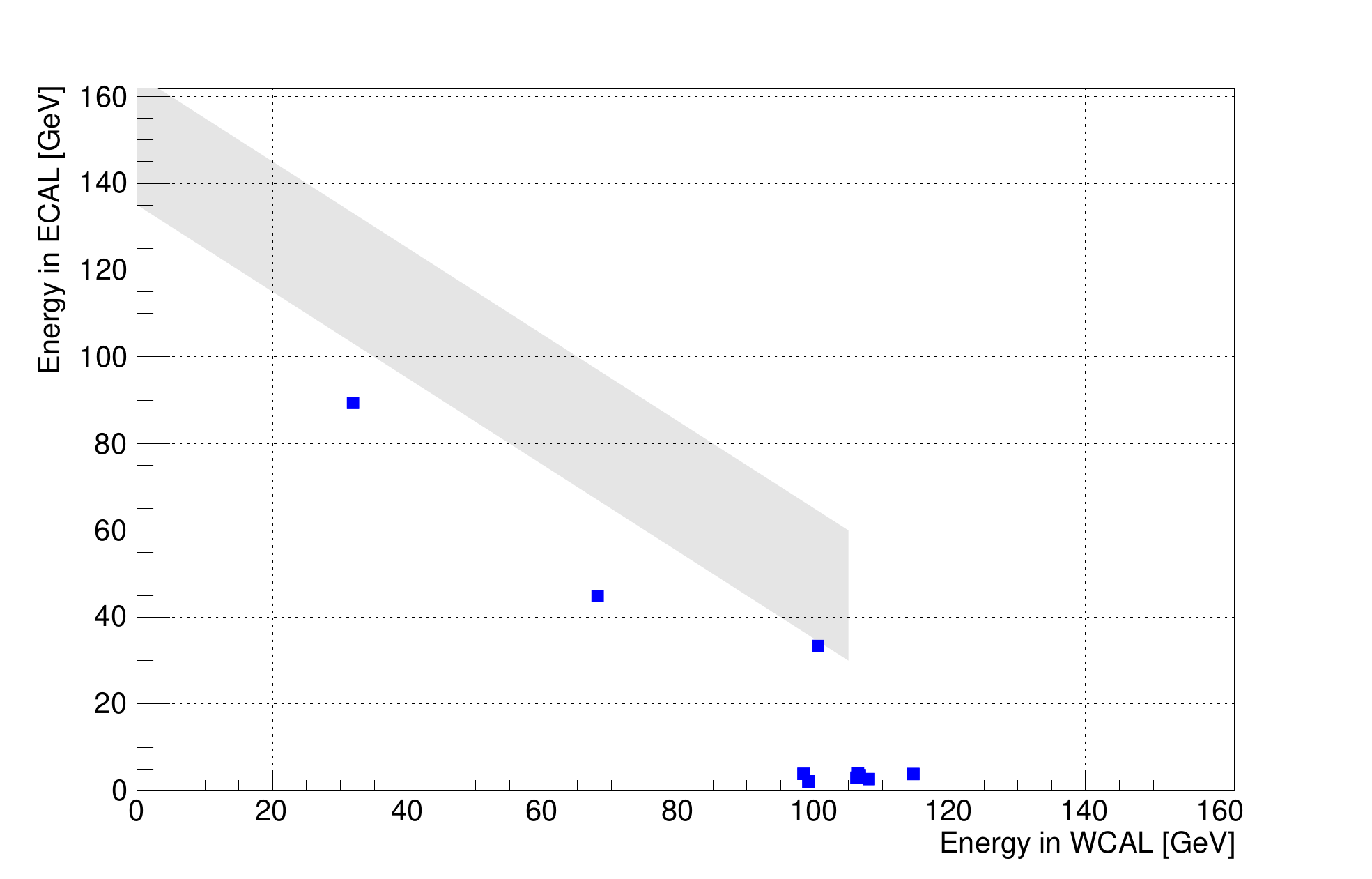}
\caption{Distribution of selected EM neutral events in the $(E_{WCAL},E_{ECAL})$ plane from the 2018 data.
Neutral events are shown as blue squares. The shadowed band represents the signal box region.}
\label{w-e}
\end{figure}

 The main background in this search comes from the $\kospio$ events from $K^0$ mainly produced by hadrons misidentified
as electrons \cite{NA64Be2017}. $K_0$ can pass the veto counters without energy deposition and decay into $\pi^0 \pi^0$. These
$\pi^0$ decay immediately into photons that can convert on the setup material into $\ee$ pair upstream of the $S_4$. The decay
chain $\kosdec$ is also possible.
We estimated this background using both simulation and data. For this, we selected the sample of neutral
events changing the cut (ii) to $E_{S_4}<0.5MIP$. This sample has 3 events in the 2017 data. No events were found with standard
criteria in the 2018 data, for this reason we relaxed for this sample the criteria iii) and vi). The distribution of neutral
events is shown in Fig.~\ref{w-e}.
The MC sample of $K^0_S$ was simulated according to distributions predicted for the hadron interactions in WCAL. With this sample
we calculated the number of neutral and signal-like events passing the criteria. This gives us the prediction of the number of background
events: 0.06 for the 2017 data and 0.005 for the 2018 data (Table~\ref{tab:BG}). The smaller number of neutral events and lower background
in the 2018 data are expected, because due to the increased distance between the WCAL and ECAL less $K^0_S$ events pass the criteria (v) and (vi).
In addition, the background is decreased due the vacuum pipe installed upstream of the $S_4$.

  The charge-exchange reaction $\pi^- p \to (\geq 1) \pi^0 + n + ...$ that can occur in the last layers of the WCAL, with decay
photons escaping the dump without interactions, accompanied by undetected secondaries, is another source of fake signal.
To evaluate this background we used the extrapolation of the charge-exchange cross sections, $\sigma \sim Z^{2/3}$, measured on
different nuclei \cite{vnb}. The beam pion flux suppression by the SRD tagging is taken into account in the estimation.
The background from the punchthrough $\pi^-$ can appear because of small inefficiency of the veto counter, mainly due to pile-up. It was estimated
using simulation and the data from the calibration runs with a hadron beam. The contribution from the beam kaon decays
in-flight $K^-  \to e^- \nu \pp (K_{e4})$ was estimated from the simulation with biased lifetime and found to be negligible.
The background from the dimuon production in the dump $e^- Z \to e^- Z \mu^+ \mu^-$ with either $\pp$ or $\mu^+ \mu^-$ pairs misidentified
as EM event in the ECAL was also found to be negligible.

\begin{table}[tbh!]
\begin{center}
\caption{Expected numbers of background events in the signal box that passed the selection criteria (i)-(vi)} \label{tab:BG}
\vspace{0.15cm}
\begin{tabular}{|c|c|c|}
\hline
\hline
Source of background                                                              & 2017 data          & 2018 data         \\
\hline
$K^0_S\to 2\pi^0$                                                                 & $0.06\pm0.034$     & $0.005\pm0.003$   \\
$\pi N\to (\geq 1) \pi^0 +n+...$                                                  & $0.01\pm0.004$     & $0.001\pm0.0004$  \\
punchthrough $\pi^-$                                                              & $0.0015\pm0.0008$  & $0.0007\pm0.0004$ \\
punchthrough $\g$                                                                 & $<0.001$           & $<0.0005$         \\
$\pi, K \to e \nu$, $K_{e4}$ decays                                               & \multicolumn{2}{c}{$ < 0.001$}         \vline \\
$eZ\to eZ \mm; \mu^\pm \to e^\pm \nu \bar\nu$                                         & \multicolumn{2}{c}{$ < 0.001$}         \vline \\
\hline
Total                                                                             & $0.07\pm 0.035$    & $0.006 \pm 0.003$ \\
\hline
\hline
\end{tabular}
\end{center}
\end{table}

\begin{figure}[tbh!!] 
\begin{center}
\includegraphics[width=0.5\textwidth]{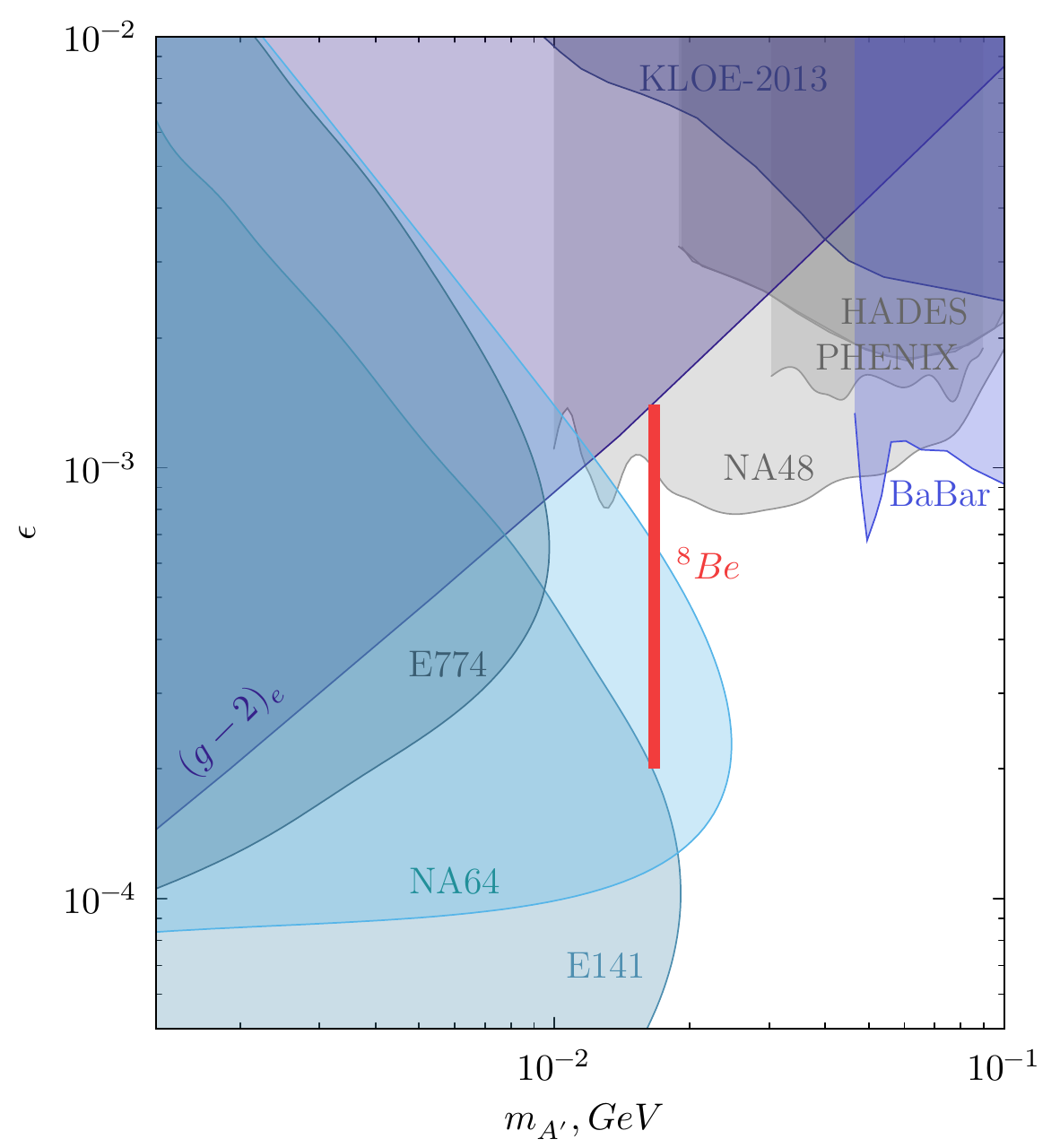}
\caption{The  90\% C.L.\ exclusion areas  in the ($m_{X}; \epsilon$) plane from the NA64 experiment (blue area). For the mass of 16.7~MeV, 
 the  $X-e^-$ coupling region excluded by NA64  is  $1.2\times 10^{-4}< \epsilon_e < 6.8~\times~10^{-4}$. The full allowed range of $\epsilon_e$ 
 explaining the $^8$Be* anomaly, $2.0\times 10^{-4} \lesssim \epsilon_e \lesssim 1.4 \times 10^{-3}$ \cite{feng1, feng2}, is also shown (red area).
 The  constraints  on the mixing $\epsilon$ from the experiments E774~\cite{bross}, E141~\cite{rio}, BaBar ~\cite{babar1}, KLOE~\cite{kloe2},
 HADES~\cite{hades}, PHENIX~\cite{phenix}, NA48~\cite{na48}, and bounds from the electron anomalous magnetic moment $(g-2)_e$ \cite{hd} are also shown.}
 \label{exclvis}
\end{center}
\end{figure}

 \par Table~\ref{tab:BG} summarizes the estimated background inside the signal box. The main part of the total background uncertainty
comes from the statistical error of the number of observed EM neutral events.
There is also the uncertainty from the cross sections of the $\pi,K$ charge-exchange reactions on heavy nuclei (30\%).

 \par After determining and optimizing the selection criteria and estimating the background levels, we examined the signal box
and found no candidates.

 \par The combined 90\% confidence level (C.L.) upper limits for the mixing strength $\epsilon$ were determined from the 90\% C.L.
upper limit for the expected number of signal events, $N_{A'}^{90\%}$ by using the modified frequentist approach for confidence levels
(C.L.), taking the profile likelihood as a test statistic in the asymptotic approximation \cite{junk,limit,Read:2002hq}. The total
number of expected signal events in the signal box was the sum of expected events from the 2017 and 2018 runs:

\begin{equation}
\Na = \sum_{i=1}^{2} N_{A'}^i = \sum_{i=1}^{2} n_{EOT}^i  P_{tot}^i n_{A'}^i(\epsilon,\ma),
\label{nev}
\end{equation}

\noindent where $n_{EOT}^i$ is the effective number of $EOT$ in run-$i$ ($5.4\times 10^{10}$ and $3\times 10^{10}$), $P_{tot}^i$ is
the signal efficiency in the run $i$, and  $n_{A'}^i(\epsilon,\ma)$ is the number of the $\aee$
decays in the decay volume with energy $E_{A'} > 25$ GeV per EOT, calculated under assumption that this decay mode is predominant,
see, e.g., Eq.(3.7) in Ref. \cite{Andreas:2013lya}.
The value $n_{EOT}^i$ takes into account the data acquisition system (DAQ) dead time.
Each $i$-th entry in this sum was calculated by simulating signal events for the corresponding beam running conditions and processing
them through the reconstruction program with the same selection criteria and efficiency corrections as for the data sample from the run-$i$.
In the overall signal efficiency for each run the acceptance loss due to pileup in the veto detectors was taken into account

The $A'$ yield from the dump was calculated as described in Ref.\cite{DMsimulation}. These calculations were cross-checked with the
calculations of Ref.\cite{Liu:2016mqv, Liu:2017htz}.
The $\lesssim 10\%$ difference between the two calculations was accounted for as a systematic uncertainty in $n_{A'}(\epsilon, \ma)$.
The total systematic uncertainty on $N_{A'}$ calculated by combining all uncertainties did not exeed $\simeq 25\%$ for all runs.
The combined 90\% C.L. exclusion limits on the mixing strength $\epsilon$ as a function of the $A'$ mass is shown in Fig.~\ref{exclvis} together 
with the constraints from other experiments. Our results exclude $X$-boson as an explanation for the $^8$Be* anomaly
for the $X-e^-$ coupling $\epsilon_e \lesssim 6.8 \times 10^{-4}$ and mass value of 16.7 MeV, leaving some unexplored region at this
mass as an interesting prospect for further searches.


We gratefully acknowledge the support of the CERN management and staff and the technical staffs of the participating institutions
for their vital contributions. This work was supported by the HISKP, University of Bonn (Germany), JINR (Dubna), MON and RAS (Russia),
ETH Zurich and SNSF Grant No. 169133 (Switzerland), and grants FONDECYT 1191103, 1190845, and 3170852, UTFSM PI M 18 13 and Basal FB0821 CONICYT (Chile).

\end{document}